\documentclass[showpacs,aps,prc,preprint,nofootinbib,superscriptaddress]{revtex4}

\usepackage{amsmath}
\usepackage{amssymb}
\def\bm{\boldsymbol}
\newcommand{\bea}{\begin{eqnarray}}
\newcommand{\eea}{\end{eqnarray}}
\newcommand{\be}{\begin{eqnarray}}
\newcommand{\ee}{\end{eqnarray}}
\newcommand{\no}{\nonumber \\}

\newcommand{\del}{\partial}

\def\vp{{\bm p}}
\def\vq{{\bm q}}

\def\vx{{\bm x}}
\def\vy{{\bm y}}

\def\vr{{\bm r}}
\def\vs{{\bm\sigma}}
\def\la{\langle}
\def\ra{\rangle}

\newcommand{\sixjsymbol}[6]{\left\{\begin{tabular}{ccc} {$#1$}&{$#2$}&{$#3$}\\
                             {$#4$}&{$#5$}&{$#6$} \end{tabular}\right\}}

\begin{document}


\title{Parity violation in low energy neutron deuteron scattering}


\author{Young-Ho Song}
\email[]{song25@mailbox.sc.edu}
\affiliation{Department of Physics and Astronomy, University of South Carolina, Columbia, SC, 29208}

\author{Rimantas Lazauskas}
\email[]{rimantas.lazauskas@ires.in2p3.fr}
\affiliation{IPHC, IN2P3-CNRS/Universit\'e Louis Pasteur BP 28,
F-67037 Strasbourg Cedex 2, France}

\author{Vladimir Gudkov}
\email[]{gudkov@sc.edu}
\affiliation{Department of Physics and Astronomy, University of South Carolina, Columbia, SC, 29208}




\date{\today}

\begin{abstract}
Parity violating effects  for low energy
elastic neutron deuteron scattering
are calculated for DDH and EFT-type of weak potentials in a Distorted Wave Born Approximation,
using realistic hadronic strong interaction wave functions, obtained by solving
three-body Faddeev equations in configuration space.
The results of relation between physical observables
and low energy constants can be used to fix
low energy constants from experiments.
Potential model dependencies of parity violating effects
are discussed.
\end{abstract}


\maketitle

\section{Introduction}

The study of parity violating (PV)  effects in low energy physics is a very sensitive tool to test  methods of calculations both of  weak and strong interactions in the Standard model. This also can be a way to search for a possible manifestations of new physics resulted in deviations from unambiguous and precise calculations of PV effects and experimental measurements. However, to use this approach, it is crucial to prove that implemented theoretical techniques  are sufficient to describe experimental data with high accuracy which exceeds experimental accuracy.
There is a large amount of experimental data for different PV effects in nuclear physics, each of which  in general agrees with theoretical predictions.
However,  in the last years it became clear (see, for example \cite{Zhu:2004vw,HolsteinUSC, DesplanqueUSC,RamseyMusolf:2006dz}  and references therein) that the traditional DDH \cite{Desplanques:1979hn} method for calculation of PV effects cannot reliably describe the whole available set of experimental data within the same set of parameters.
If this is not the  manifestation of new physics, which is very unlikely for the current accuracy of experimental measurements and theoretical calculations, then this discrepancy could be blamed on systematic errors in experimental data, theoretical uncertainties in calculations of strong interactions at low energy, or
 it might be that DDH approach is not adequate for the description of the set of precise experimental data because it is based on a number of models and assumptions.
To resolve this  discrepancy and to eliminate  nuclear model dependent factors in calculations, it is necessary to focus on the analysis of new and existing experimental data for different PV parameters in  few-body systems, where calculations of nuclear related effects can be done with a high precision.
Recently  new approach, based on the effective field theory (EFT), has been introduced for a model independent parametrization of PV effects (see, papers \cite{Zhu:2004vw,RamseyMusolf:2006dz} and references therein), and some calculations for two-body systems have been done \cite{Liu:2006dm}.
The power of the EFT approach for parametrization of all PV effects in terms of a small number of constants could be utilized if we can analyze a large enough number of PV effects to be able to constrain all free parameters of the theory which are usually called  low energy constants (LEC). Thus, one can guarantee the adequate description (parametrization) of the strong interaction hadronic parts and weak interaction constants for symmetry violating observables.
Unfortunately, the number of experimentally measured (and independent in terms of unknown LECs) PV effects in two body systems is not enough to constrain all LECs. In spite of the fact that five independent observable parameters in a two body system could fix  five unknown PV LECs \cite{Girlanda:2008ts,Phillips:2008hn,Shin:2009hi,Schindler:2009wd},
 it is impossible to measure all of them using existing experimental techniques.
Therefore, one has to include into analysis few-body systems and even heavier nuclei, the latter of which are actually preferable from the experimental point of view, because as a rule,  the measured effects in nuclei are much larger than in nucleon-nucleon system due to nuclear enhancement factors \cite{Sushkov:1982fa,Bunakov:1982is,Gudkov:1991qg}.

The natural and unambiguous  way to verify the applicability of the EFT for the calculation of symmetry violating effects in  nuclear reactions requires a development of a regular and self consistent approach for calculation of PV amplitudes in three-body (few-body) systems \cite{Gudkov:2010pt}, with a hope to extend the formalism for the description of many body systems.
This systematic approach for the solution of three-body PV scattering problem in EFT framework \cite{Gudkov:2010pt} requires additional numerical efforts and will be presented elsewhere.
As a first step for the clarification of the possible difference in contributions to PV effects from DDH and EFT-type potentials,
one can use  a ``hybrid'' method (similar to the method used in paper \cite{Schiavilla:2008ic}) for the simplest process of neutron-deuteron scattering.   We calculate three-body wave functions with realistic Hamiltonians of strong interaction
using exact  Faddeev equations in configuration space, and then, calculate PV effects in the first order of perturbation   with DDH potential and potentials derived in EFT formalism. In the next section, we present our formalism for the calculation PV effects for elastic neutron-deuteron scattering with different set of nucleon weak potentials, with DDH and weak potentials obtained from pionless and pionful EFTs.
Then, we present results of numerical calculations and discussions.

\section{Formalism}
We treat weak nucleon interactions as a  perturbation and calculate three-body  wave functions exactly using Faddeev equations with  phenomenological potentials for strong interactions.
Similar hybrid approach has been successfully applied
to the weak and electromagnetic processes involving
three-body and four-body hadronic systems \cite{Song:2007bj,Song:2008zf,Lazauskas:2009nw,
Park:2002yp,Pastore:2009is,Girlanda:2010vm}.
We consider three types of parity violating potentials.
The first one is the standard DDH potential which is based on
meson exchange mechanism of nucleon-nucleon interactions.
The second and third potentials are derived from
pionless and pionful versions of effective field theory
with parity violating hadronic interactions.
Instead of calculating parity violating amplitudes by
summing PV diagrams in EFT,
we use these potentials to calculate PV effects. This is a simplification, which we call a ``hybrid'' approach.

\subsection{Observables}
\label{sec:observables}
Since PV effects in neutron-deuteron system are  very small, we consider only coherent processes which are related to the propagation of neutrons through unpolarized deuteron target and, therefore, do not have an additional suppression in low energy region.  Then, two PV observable parameters are the angle $\phi$ of rotation of neutron polarization around neutron momentum and the relative difference of total  cross sections $P=(\sigma_+-\sigma_-)/(\sigma_++\sigma_-)$ for neutrons with opposite helicities. The value of the angle of neutron spin rotation per unit length of the target sample can be expressed in terms of elastic scattering amplitudes at zero angle for opposite helicities $f_+$ and $f_-$ as
\begin{equation}
\label{eq:neutronspinrot1}
\frac{d\phi}{dz}
=-\frac{2\pi N}{p}{\rm Re}
  \left(f_{+}
  -f_{-}\right),
\end{equation}
where $N$ is a number of target nuclei per unit volume and
$p$ is a relative neutron momentum. Using optical theorem, one can write the relative difference of total cross sections $P$ in terms of these amplitudes as
\bea
P=\frac{{\rm Im}\left(f_{+}
  -f_{-}
\right)}
{{\rm Im}\left(f_{+}
  +f_{-}
\right)}.
\eea

It is convenient to represent the amplitudes in terms of matrix $\hat{R}$ which is related to
scattering matrix $\hat{S}$ as $\hat{R}=\hat{1}-\hat{S}$.
With partial waves decomposition
for the case of neutron-deuteron scattering

\bea
|\vp, m_n,m_d\ra=\sum_{l_y l_y^z}\sum_{SM,JJ^z}
 |p,(l_y {\cal S}) J J^z\ra
 \la J J^z|l_y l_y^z, {\cal S}M\ra
 \la {\cal S} M|\frac{1}{2}m_n,1 m_d\ra
 Y^*_{l_yl_y^z}(\hat{p}),
\eea
where $l_y$ is an orbital angular momentum
between neutron and deuteron,  ${\cal S}$ is a sum of
neutron spin and deuteron total angular momentum, and $J$ is the total angular momentum of
the neutron-deuteron system,
the above equations can be written at low energies as
\bea
\label{eq:neutronspinrot02}
\frac{1}{N}\frac{d\phi}{dz}&=&\frac{2\pi}{9 p^2}\mbox{Im}
  \left[ R^{\frac{1}{2}}_{1\frac{1}{2},0\frac{1}{2}}
        +R^{\frac{1}{2}}_{0\frac{1}{2},1\frac{1}{2}}
   -2\sqrt{2} R^{\frac{1}{2}}_{1\frac{3}{2},0\frac{1}{2}}
   -2\sqrt{2} R^{\frac{1}{2}}_{0\frac{1}{2},1\frac{3}{2}}
  \right. \no & &  \left.
  +4R^{\frac{3}{2}}_{1\frac{1}{2},0\frac{3}{2}}
  +4R^{\frac{3}{2}}_{0\frac{3}{2},1\frac{1}{2}}
     -2\sqrt{5} R^{\frac{3}{2}}_{1\frac{3}{2},0\frac{3}{2}}
     -2\sqrt{5} R^{\frac{3}{2}}_{0\frac{3}{2},1\frac{3}{2}}
\right]
\eea
and
\bea
\label{eq:P2}
P&=&\frac{1}{3}{\rm Re}
    \left[ R^{\frac{1}{2}}_{1\frac{1}{2},0\frac{1}{2}}
        +R^{\frac{1}{2}}_{0\frac{1}{2},1\frac{1}{2}}
   -2\sqrt{2} R^{\frac{1}{2}}_{1\frac{3}{2},0\frac{1}{2}}
   -2\sqrt{2} R^{\frac{1}{2}}_{0\frac{1}{2},1\frac{3}{2}}
    +4R^{\frac{3}{2}}_{1\frac{1}{2},0\frac{3}{2}}
  \right. \no & &  \left.
   +4R^{\frac{3}{2}}_{0\frac{3}{2},1\frac{1}{2}}
     -2\sqrt{5} R^{\frac{3}{2}}_{1\frac{3}{2},0\frac{3}{2}}
     -2\sqrt{5} R^{\frac{3}{2}}_{0\frac{3}{2},1\frac{3}{2}}
\right]/
     {\rm Re} \left[ R^{\frac{1}{2}}_{0\frac{1}{2},0\frac{1}{2}}
              +2 R^{\frac{3}{2}}_{0\frac{3}{2},0\frac{3}{2}}
             \right]       ,
\eea
where
$R^{J}_{l^\prime {\cal S}^\prime  ,l {\cal S}}=\la l^\prime {\cal S}^\prime|R^{J}|l {\cal S}\ra$,  unprimed and  primed parameters correspond to initial and final states.
Since we are interested in low energy neutron scattering,  it would be sufficient to include only $s$- and
$p$-waves contributions to parity violating amplitudes; for the total cross section (the denominator in the last equation), we keep only dominant contributions from $s$-wave neutrons. It should be noted that time-reversal invariance leads to the relation $\la 1 {\cal S}^\prime|R^{J}|0 {\cal S}\ra=\la 0 {\cal S}^\prime|R^{J}|1 {\cal S}^\prime\ra$ between matrix elements, therefore, only half of parity violating amplitudes are independent.

Nucleon-nucleon interaction can be written as a sum $V=V_{pc}+V_{pv}$ of the parity conserving ($V_{pc}$) and weak parity violating ($V_{pv}$) terms.
Due to the weakness of parity violating  interaction,  one can  use Distorted Wave Born Approximation (DWBA) to
calculate PV amplitudes with a high level of accuracy as
\begin{equation}
R^J_{l'_y {\cal S}',l_y{\cal S}}
\simeq 4 i^{-l'_y+l_y+1} \mu p \;
{}^{(-)}_{pc} \la \Psi,(l'_y{\cal S}')J J^z|V_{pv}|\Psi,(l_y{\cal S})J J^z\ra^{(+)}_{pc},
\end{equation}
where  $\mu$ is a neutron-deuteron
reduced mass and  $|\Psi,(l'_y{\cal S}')J J^z\rangle^{(\pm )}_{pc}$
are  solutions of  3-body Faddeev equations in configuration space
for parity conserving strong interaction Hamiltonian, defined by $V_{PC}$ and normalized as
described in section~\ref{sec:FE}.
The factor $i^{-l'_y+l_y}$ in this expression
is introduced to match the $R$-matrix definition
in the modified spherical harmonics convention \cite{Varshalovich}
with the  wave functions which are calculated in this paper using spherical harmonics
convention.

In the rest of the paper, we use only wave functions calculated for parity
conserving potentials and, therefore, will omit subscript $PC$.

As will be explained in section~\ref{sec:FE}, we use
jj-coupling scheme (with a basis states $|l_y j_y\rangle$) when solving Faddeev equtions.
One can transform $jj$-basis states into $l_y {\cal S}$-basis  by means of
\bea
|[l_y\otimes(s_k \otimes j_x)_{\cal S}]_{J J_z}\rangle
&=&\sum_{j_y}|[j_x\otimes (l_y \otimes s_k)_{j_y}]_{J J_z}\rangle\no
& & \times (-1)^{j_x+j_y-J}(-1)^{l_y+s_k+j_x+J}
 [(2 j_y+1)(2 {\cal S}+1)]^{\frac{1}{2}}
 \sixjsymbol{l_y}{s_k}{j_y}{j_x}{J}{{\cal S}},
\eea
One interesting observation is that the neutron spin rotation, as well as
parameter P,
in $|l_y j_y\rangle$ basis involves
potential matrix elements only between $j_y=\frac{1}{2}$ states.

It should be noted that  at low energy the ${\rm Im}(R^J_{l'_y {\cal S}',l_y{\cal S}})\sim p^{l'_y+l_y+1}$, and thus
the expression~eq.(\ref{eq:neutronspinrot02}) for the angle $\phi$ of neutron spin rotation is finite and well defined
in the zero energy limit of the n-d scattering. Numerically, it is calculated by evaluating expression
${\rm Im}(R^J_{l'_y {\cal S}',l_y{\cal S}})/p^{l'_y+l_y+1}$ at zero energy.
On the other hand, ${\rm Re}(R^J_{l'_y {\cal S}',l_y{\cal S}})\sim p\cdot {\rm Im}(R^J_{l'_y {\cal S}',l_y{\cal S}})$ at
low energy, and thus the real part of this quantity vanishes
in the zero energy limit.
Therefore, the parameter $P$ is calculated
at $15$ KeV neutron kinetic energy in the laboratory system,
where both imaginary and real parts of the R-matrix elements
become comparable in magnitude and thus can be discerned
numerically.

\subsection{The parity violating potentials}
To understand the possible difference in the description of  parity violating effects by DDH and EFT-type for potentials, we  compare calculations with the DDH potential\cite{Desplanques:1979hn} and two different choices of EFT potentials:  the potential derived from pionless EFT lagrangian \cite{Zhu:2004vw} and the potential derived from pionful EFT Lagrangian \cite{Zhu:2004vw}.
It was shown \cite{Schiavilla:2008ic} that all these three potentials can be expanded in terms of a set of $O^{(n)}_{ij}$ operators   as
\bea
v_{ij}^\alpha=\sum_{n} c_n^\alpha O^{(n)}_{ij},\quad
\mbox{$\alpha=$ DDH or pionless EFT or pionful EFT}
\eea
with  parameters $c_n^\alpha$
and operators $O^{(n)}_{ij}$
given in the Table \ref{tbl:pvpotential}.

\begin{table}
\caption{\label{tbl:pvpotential}
 Parameters and operators of parity violating potentials.
 $\pi NN$ coupling $g_{\pi NN}$ can be represented by $g_A$ by using
Goldberger-Treiman relation, $g_\pi=g_A m_N/ F_\pi$ with $F_\pi=92.4$ MeV.
${\cal T}_{ij}\equiv (3\tau_i^z\tau_j^z-\tau_i\cdot\tau_j)$.
Scalar function
$\tilde{L}_\Lambda(r)\equiv 3L_\Lambda(r)-H_\Lambda(r) $.
}
\begin{ruledtabular}
\begin{tabular}{cccccccc}
$n$ & $c_n^{DDH}$ & $f_n^{DDH}(r)$ & $c_n^{\not{\pi}}$ & $f_n^{\not{\pi}}(r)$ & $c_n^{\pi}$ & $f_n^{\pi}(r)$ & $O^{(n)}_{ij}$ \\
\hline
$1$ & $+\frac{g_\pi }{2\sqrt{2} m_N}h_\pi^1$ & $f_\pi(r)$ &
      $\frac{2\mu^2}{\Lambda^3_\chi}   C^{\not{\pi}}_6$ &
      $f^{\not{\pi}}_\mu(r)$ &
      $+\frac{g_\pi }{2\sqrt{2} m_N}h_\pi^1$ & $f_\pi(r)$ &
      $(\tau_i\times\tau_j)^z(\vs_i+\vs_j)\cdot{\bm X}^{(1)}_{ij,-}$
\\
$2 $ & $ -\frac{g_\rho}{m_N}h_\rho^0 $ & $ f_\rho(r) $ & $
      0 $ & $ 0 $ & $
      0 $ & $ 0 $ & $
      (\tau_i\cdot\tau_j)(\vs_i-\vs_j)\cdot{\bm X}^{(2)}_{ij,+}$
\\
$3 $ & $ -\frac{g_\rho(1+\kappa_\rho)}{m_N} h_\rho^0 $ & $f_\rho(r) $ & $
      0 $ & $ 0$ & $
       0 $ & $ 0 $ & $
      (\tau_i\cdot\tau_j)(\vs_i\times\vs_j)\cdot{\bm X}^{(3)}_{ij,-}$
\\
$4 $ & $ -\frac{g_\rho}{2 m_N} h_\rho^1 $ & $ f_\rho(r) $ & $
      \frac{\mu^2}{\Lambda^3_\chi}(C^{\not{\pi}}_2+C^{\not{\pi}}_4)
      $ & $ f_\mu^{\not{\pi}}(r) $ & $
      \frac{\Lambda^2}{\Lambda^3_\chi}(C^{\pi}_2+C^{{\pi}}_4) $ & $
      f_\Lambda(r) $ & $
      (\tau_i+\tau_j)^z(\vs_i-\vs_j)\cdot{\bm X}^{(4)}_{ij,+}$
\\
$5  $ & $  -\frac{g_\rho(1+\kappa_\rho)}{2 m_N}h_\rho^1 $ & $ f_\rho(r)
     $ & $ 0 $ & $ 0 $ & $
     \frac{2\sqrt{2}\pi g_A^3\Lambda^2}{\Lambda_\chi^3}h^1_\pi $ & $
      L_\Lambda(r) $ & $
      (\tau_i+\tau_j)^z(\vs_i\times\vs_j)\cdot{\bm X}^{(5)}_{ij,-}$
\\
$6 $ & $ -\frac{g_\rho}{2\sqrt{6} m_N}h_\rho^2 $ & $ f_\rho(r) $ & $
  -\frac{2\mu^2}{\Lambda^3_\chi}C_5^{\not{\pi}} $
  & $   f^{\not{\pi}}_\mu(r)$ & $
  -\frac{2\Lambda^2}{\Lambda^3_\chi}C_5^{{\pi}} $
  & $ f_\Lambda(r)$ & $
   {\cal T}_{ij}
   (\vs_i-\vs_j)\cdot{\bm X}^{(6)}_{ij,+}$
\\
$7 $ & $ -\frac{g_\rho(1+\kappa_\rho)}{2\sqrt{6} m_N}h_\rho^2 $ & $ f_\rho(r) $ & $
   0 $ & $ 0 $ & $
    0  $ & $ 0 $ & $
   {\cal T}_{ij}(\vs_i\times\vs_j)\cdot{\bm X}^{(7)}_{ij,-}$
\\
$8 $ & $ -\frac{g_\omega}{m_N}h_\omega^0 $ & $ f_\omega(r) $ & $
   \frac{2\mu^2}{\Lambda^3_\chi} C_1^{\not{\pi}} $ & $ f_\mu^{\not{\pi}}(r) $ & $
   \frac{2\Lambda^2}{\Lambda^3_\chi} C_1^{{\pi}} $ & $ f_\Lambda(r) $ & $
      (\vs_i-\vs_j)\cdot{\bm X}^{(8)}_{ij,+}$
\\
$9  $ & $  -\frac{g_\omega(1+\kappa_\omega)}{m_N} h_\omega^0 $ & $ f_\omega(r) $ & $
   \frac{2\mu^2}{\Lambda_\chi^3}\tilde{C}^{\not{\pi}}_1 $ & $ f_\mu^{\not{\pi}}(r) $ & $
   \frac{2\Lambda^2}{\Lambda_\chi^3}\tilde{C}^{{\pi}}_1 $ & $ f_\Lambda(r) $ & $
   (\vs_i\times\vs_j)\cdot{\bm X}^{(9)}_{ij,-}$
\\
$10 $ & $ -\frac{g_\omega}{2 m_N} h_\omega^1 $ & $ f_\omega(r) $ & $
     0 $ & $ 0 $ & $
      0  $ & $ 0  $ & $
     (\tau_i+\tau_j)^z(\vs_i-\vs_j)\cdot{\bm X}^{(10)}_{ij,+}$
\\
$11 $ & $ -\frac{g_\omega(1+\kappa_\omega)}{2m_N} h^1_\omega $ & $ f_\omega(r) $ & $
    0 $ & $  0 $ & $
     0  $ & $  0  $ & $
    (\tau_i+\tau_j)^z(\vs_i\times\vs_j)\cdot{\bm X}^{(11)}_{ij,-}$
\\
$12 $ & $ -\frac{g_\omega h_\omega^1-g_\rho h_\rho^1}{2m_N} $ & $ f_\rho(r) $ & $
   0 $ & $ 0 $ & $
    0 $ & $ 0   $ & $
   (\tau_i-\tau_j)^z(\vs_i+\vs_j)\cdot{\bm X}^{(12)}_{ij,+}$
\\
$13 $ & $ -\frac{g_\rho}{2m_N} h^{'1}_\rho $ & $ f_\rho(r) $ & $
      0 $ & $ 0 $ & $
       -\frac{\sqrt{2}\pi g_A\Lambda^2}{\Lambda_\chi^3} h_\pi^1 $ & $ L_\Lambda(r) $ & $
     (\tau_i\times\tau_j)^z(\vs_i+\vs_j)\cdot{\bm X}^{(13)}_{ij,-}$
\\
$14$ & 0  & 0 &  0 & 0 &
 $\frac{2\Lambda^2}{\Lambda^3_\chi}   C^{{\pi}}_6$ &
 $f_\Lambda(r)$ &
 $(\tau_i\times\tau_j)^z(\vs_i+\vs_j)\cdot{\bm X}^{(14)}_{ij,-}$
\\
$15$   &  0  &  0   & 0 &  0   &
   $\frac{\sqrt{2}\pi g^3_A\Lambda^2}{\Lambda_\chi^3} h_\pi^1 $ &
   $\tilde{L}_\Lambda(r)$ &
   $(\tau_i\times\tau_j)^z(\vs_i+\vs_j)\cdot{\bm X}^{(15)}_{ij,-}$
\end{tabular}
\end{ruledtabular}
\end{table}

One can see that operators $O^{(n)}_{ij}$ are  products
 of isospin, spin,  and
vector operators ${\bm X}^{(n)}_{ij,\pm}$ defined as
\bea
{\bm X}^{(n)}_{ij,+} & \equiv&[\vp_{ij},f_n(r_{ij})]_{+},\no
{\bm X}^{(n)}_{ij,-}&\equiv&i [\vp_{ij},f_n(r_{ij})]_{-},
\eea
where $\vp_{ij}\equiv\frac{(\vp_i-\vp_j)}{2}$.

For the DDH potential, radial functions $f_{x}(r)$, $x=\pi,\rho$, and $\omega$ are modified Yukawa functions,
\bea
f_{x}(r)=\frac{1}{4\pi r}\left\{e^{-m_x r}-e^{-\Lambda_x r}\left[
                          1+\frac{\Lambda_x r}{2}\left(1-\frac{m_x^2}{\Lambda_x^2}\right)
                          \right]\right\}.
\eea

For pionless EFT  ($\not{\pi}$EFT) one,
$f_n(r)$ are described by single function $f_\mu(r)$,
\bea
f_{\mu}(r)=\frac{1}{4\pi r}e^{-\mu r},
\eea
with $\mu\simeq m_\pi$.

For the case of pionful EFT model ($\pi$EFT), there are long range interactions
from one pion exchange($V_{-1,LR}$) and from their corrections ($V_{1,LR}$), middle range interactions
due to two pion exchange ($V_{1,MR}$), and short
range interactions ($V_{1,SR}$)
due to nucleon contact terms.
The radial part of the leading term of long range one pion exchange, $V_{-1,LR}$ ,
is described by the function $f_\pi(r)$.
Since one-pion exchange contribution
is dominated by long range part, we do not use a
regulator for it, i.e. we assume that the
long range interactions have the same radial functions  $f_\pi(r)$ as  DDH
potential with infinite cutoff.
The short range interaction $V_{1,SR}$ in pionful theory
has the same structure as for pionless EFT;
however,  in spite of the structural similarity,
their meanings are rather different.
One can ignore the higher order corrections of long range interactions, $V_{1,LR}$,
 because they can either be absorbed by renormalization of low energy constants \cite{Liu:2006dm} or  suppressed.
The middle range interactions $V_{1,MR}$
are described by functions $L(q)$ and $H(q)$ in momentum space
\bea
L(q)\equiv \frac{\sqrt{4 m_\pi^2+\vq^2}}{|\vq|}\ln\left(
                         \frac{\sqrt{4 m_\pi^2+\vq^2}+|\vq|}{2m_\pi}\right),
\quad
H(q)\equiv \frac{4 m_\pi^2}{4 m_\pi^2+\vq^2}L(q),
\eea
where, $q^\mu=(q^0,\vq)=p_1^\mu-p^{'\mu}_1=p^{'\mu}_2-p_2^\mu$.
To calculate two pion exchange functions
(divergent at large $q$) in spacial representation
, we use regulators
$\frac{(\Lambda^2-4m_\pi^2)^2}{(\Lambda^2+\vq^2)^2}$.
For the sake of simplicity, we use only one cutoff parameter
with  the same regulator, both for middle range and for short range interactions.  Then, one can write
\bea
\{L_\Lambda(r),H_\Lambda(r),f_\Lambda(r)\}
=\frac{1}{\Lambda^2}
\int \frac{d^3 q}{(2\pi)^3} e^{-i\vq\cdot\vr}
\frac{(\Lambda^2-4m_\pi^2)^2}{(\Lambda^2+\vq^2)^2} \{L(q),H(q),1\}.
\eea
In the given representation,
coefficients $c_n^\alpha$ have $\mbox{fm}$ dimension
and scalar functions $f_n^\alpha(r)$ have $\mbox{fm}^{-1}$
dimension.
One can see that only the
new operator structure, which is not included
in DDH or pionless EFT, is due to $V^{PV}_{1,LR}$.
Therefore, pionful EFT does not introduce
new operator structure, provided
we neglect $V^{PV}_{1,LR}$ term \cite{Liu:2006dm,Hyun:2006mp}.

To see a sensitivity to the choice of cutoffs for  parity violating
potentials, we used two set of cutoff parameters
for each models,
which are listed in the Table \ref{tbl:pv:parameters}.

Using the discussed above three potentials, one can represent parity violating amplitudes as a linear expansion in terms of given set of matrix elements for corresponding operators $O^{(n)}_{ij}$.
Thus, the angle of neutron spin rotation
can be written as
\bea
\frac{1}{N}\frac{d\phi}{dz}
=\sum_{n=1}^{13} c_n^\alpha I_n^\alpha,
\eea
and the parameter $P$ as
\bea
P=\sum_{n=1}^{13} {c}_n^\alpha {\tilde I}_n^\alpha,
\eea
 in terms of
coefficients $I_n^{\alpha}$ and $\tilde{I}_n^{\alpha}$
with $\alpha=\mbox{DDH-I,II},\ \not{\pi}\mbox{EFT-I,II},\
{\pi}\mbox{EFT-I,II}$ for different potentials
and cutoff parameters.

\begin{table}
\caption{\label{tbl:pv:parameters}
Parameter of parity violating potentials
in GeV units.
We used masses of mesons
$m_\pi$, $m_\rho$, and $m_\omega$, respectively, as
$0.138$, $0.771$, and $0.783$ in DDH potential.
}
\begin{ruledtabular}
\begin{tabular}{cccccccc}
 & $\Lambda_\pi$ & $\Lambda_\rho$ & $\Lambda_\omega$ &  & $\mu$
 & & $\Lambda$\\
DDH-I & $1.72$ & $1.31$ & $1.50$ & $\not\pi$EFT-I & $0.138$ &
 $\pi$EFT-I & $0.8$      \\
DDH-II & $\infty$ & $\infty$ & $\infty$ & $\not\pi$EFT-II & $1.0$
 & $\pi$EFT-II & $1.0$
\end{tabular}
\end{ruledtabular}
\end{table}

\subsection{Faddeev wave function\label{sec:FE}}
To obtain 3-body wave functions for neutron-deuteron scattering with parity conserving interactions, we
solve Faddeev equations (also often called Kowalski-Noyes
equations) in configuration space \cite{Faddeev:1960su,Lazauskas:2004hq}. For isospin invariant interactions (with nucleon masses
fixed to $\hbar ^{2}/m=41.471$ MeV$\cdot$fm), three Faddeev equations
become formally identical, having the form
\begin{equation}
\left( E-H_{0}-V_{ij}\right) \psi _{k}
=V_{ij}(\psi _{i}+\psi _{j}),
\label{EQ_FE}
\end{equation}
where $(ijk)$ are particle indices, $H_{0}$ is kinetic energy operator,
$V_{ij}$ is two body force between particles $i$, and $j$,
$\psi _{k}=\psi_{ij,k}$ is
Faddeev component.

The wave function in Faddeev formalism is the sum of three Faddeev
components,
\bea
\Psi({\bm x},{\bm y})=\psi_1({\bm x}_1,{\bm y}_1)
                      +\psi_2({\bm x}_2,{\bm y}_2)
                      +\psi_3({\bm x}_3,{\bm y}_3).
\eea
Using relative Jacobi coordinates
$\vx_{k}=(\vr_{j}-\vr_{i})\smallskip $
and
$\vy_{k}=
\frac{2}{\sqrt{3}}(\vr_{k}-\frac{\vr_{i}+\vr_{j}}{2})$,
one can expand these  Faddeev components in bipolar harmonic basis:
\begin{equation}
\psi _{k}=\sum\limits_{\alpha }\frac{F_{\alpha }(x_{k},y_{k})}{x_{k}y_{k}}%
\left\vert \left( l_{x}\left( s_{i}s_{j}\right) _{s_{x}}\right)
_{j_{x}}\left( l_{y}s_{k}\right) _{j_{y}}\right\rangle _{JM}\otimes
\left\vert \left( t_{i}t_{j}\right) _{t_{x}}t_{k}\right\rangle _{TT_{z}},
\label{EQ_FA_exp}
\end{equation}%
where index $\alpha $ represents all allowed combinations of the
quantum numbers presented in the brackets: $l_{x}$ and $l_{y}$ are the
partial angular momenta associated with respective Jacobi coordinates, $%
s_{i} $ and $t_{i}$ are the spins and isospins of the individual particles. Functions
$F_{\alpha }(x_{k},y_{k})$ are called partial Faddeev amplitudes.
It should be noted that the total angular momentum $J$ as well as its
projection $M$ are conserved, but the total isospin $T$ of the system is not conserved due to the presence of charge dependent terms in nuclear
interactions.

 Boundary conditions for Eq.~(\ref{EQ_FE}) can be written in the
Dirichlet form. Thus, Faddeev amplitudes satisfy the regularity conditions:
\begin{equation}
F_{\alpha }(0,y_{k})=F_{\alpha }(x_{k},0)=0.  \label{BC_xyz_0}
\end{equation}%
For neutron-deuteron scattering
with energies below the break-up threshold, Faddeev components
vanish  for $\mathbf{x}_{k}\rightarrow\infty$.
If $\mathbf{y}_{k}\rightarrow\infty $, then interactions between the particle $k$ and the cluster $ij$
are negligible, and Faddeev components $\psi_{i}$ and
$\psi_{j}$ vanish. Then, for
the component $\psi_{k}$, which describes
the plane wave of the particle $k$ with respect
to the bound particle pair $ij$,
\begin{eqnarray}
\lim_{y_k\to \infty}
\psi_{k}(\mathbf{x}_{k},\mathbf{y}_{k} )_{l_{n}j_{n}}&=&
\frac{1}{\sqrt{3}}\sum\limits_{j_{n}^{\prime }l_{n}^{\prime }}
\left| \left\{\phi_{d}(\mathbf{x}_{k})\right\}_{j_{d}}\otimes \left\{ Y_{l_{n}^{\prime }}(\mathbf{\hat{y}}_{k})\otimes s_{k}\right\}
_{j^{\prime}_{n}}\right\rangle_{JM}
\otimes \left\vert \left( t_{i}t_{j}\right)_{t_{d}}t_{k}\right\rangle_{\frac12,-\frac12}
\notag \\
&&
\times \frac{i}{2}\left[\delta_{l_{n}^{\prime}j_{n}^{%
\prime},l_{n}j_{n}}h_{l^\prime_{n}}^{-}(pr_{nd})-S_{l_{n}^{\prime}j_{n}^{%
\prime},l_{n}j_{n}}h_{l^\prime_{n}}^{+}(pr_{nd})\right],  \label{eq_as_beh}
\end{eqnarray}%
where deuteron, being formed from nucleons $i$ and $j$, has quantum numbers
$s_{d}=1$,  $j_{d}=1$, and $t_{d}=0$, and its wave function
$\phi_{d}(\mathbf{x}_{k})$ is normalized to unity. Here,
 $r_{nd}=(\sqrt{3}/2)y_{k}$ is relative distance between
neutron and deuteron target, and $h_{l_{n}}^{\pm }$ are
the spherical Hankel
functions. The expression~(\ref{eq_as_beh}) is normalized to satisfy a condition of  unit flux for
$nd$ scattering
wave function.

For the cases where Urbana type three-nucleon interaction (TNI) is included,
we modify the Faddeev equation (\ref{EQ_FE}) into
\begin{equation}
\left( E-H_{0}-V_{ij}\right) \psi_{k}=V_{ij}(\psi_{i}+\psi_{j})+\frac{1}{2}(V_{jk}^{i}+V_{ki}^{j})\Psi
\end{equation}%
by noting that the TNI among particles $ijk$ can be written as sum of three terms:
$V_{ijk}=V_{ij}^{k}+V_{jk}^{i}+V_{ki}^{j}$.

\subsection{Evaluation of matrix elements}
Due to anti-symmetry  of  the total wave function in isospin basis, one has
$\la \Psi| V_{12}+V_{23}+V_{31} |\Psi\ra
=3\la \Psi| V_{ij}| \Psi\ra$ for any pair $i\neq j$.

Using decomposition of momentum $\vp$,
\bea
\vp=-i \nabla_x=-i\left(\hat{x}\frac{\del}{\del x}
         +\frac{1}{x}\hat{\nabla}_{\Omega}\right),
\eea
we can represent general matrix elements of local two-body
parity violating potential operators as
\bea
{}^{(-)}\la \Psi_f|O|\Psi_i\ra^{(+)}
=
(\frac{\sqrt{3}}{2})^3\sum_{\alpha\beta}\left[\int dx x^2 dy y^2
 \left(\frac{\widetilde{F}^{(+)}_{f,\alpha}(x,y)}{xy}\right)
 \hat{X}(x)
 \left(\frac{\widetilde{F}^{(+)}_{i,\beta}(x,y)}{xy}\right)
 \right] \la \alpha|\hat{O}(\hat{x})|\beta\ra,
\eea
where
$(\pm)$ means outgoing and incoming boundary conditions
and $\hat{X}(x)$ is derivative of scalar function or
derivative of wave function with respect to $x$.
(Note that we have used the fact that $(\widetilde{F}^{(-)})^*=\widetilde{F}^{(+)}$.)
The partial amplitudes $\widetilde{F}_{i(f),\alpha}(x,y)$ represent
the total systems wave function in
one selected basis set among three possible angular momentum coupling
sequences for three particle angular momenta:
\begin{equation}
\Psi_{i(f)}(x,y)=\sum\limits_{\alpha }\frac{\widetilde{F}_{i(f),{\alpha }}(x,y)}{xy}%
\left\vert \left( l_{x}\left( s_{i}s_{j}\right) _{s_{x}}\right)
_{j_{x}}\left( l_{y}s_{k}\right) _{j_{y}}\right\rangle _{JM}\otimes
\left\vert \left( t_{i}t_{j}\right) _{t_{x}}t_{k}\right\rangle _{TT_{z}}.
\label{EQ_FW_exp}
\end{equation}

The ``angular'' part of the matrix element is
\bea
\la \alpha|\hat {O}(\hat{x})|\beta\ra
\equiv \int d\hat{x}\int d\hat{y}
{\cal Y}^\dagger_\alpha(\hat{x},\hat{y})\hat{O}(\hat{x})
{\cal Y}_\beta(\hat{x},\hat{y}),
\eea
where ${\cal Y}_\alpha(\hat{x},\hat{y})$
is a tensor bipolar spherical harmonic
with a quantum number $\alpha$.
One can see that operators for ``angular" matrix elements  have the following structure:
\bea
\hat{O}(\hat{x})=(\tau_2\odot\tau_3)(\vs_2\circledcirc\vs_3)\cdot \hat{x},\mbox{ or }
(\tau_2\odot\tau_3)(\vs_2\circledcirc\vs_3)\cdot \nabla_{\Omega},
\eea
where $\odot,\circledcirc=\pm,\times$.
The explicit values of these matrix elements
are summarized in the appendix.

\section{Results and Discussions}

As it was mentioned in the previous section,
because of low energy property of $R^J_{\alpha'\alpha}$,
it is convenient to present results for elements $R^J_{l'_y,l_y}$  in terms of a ratio,
\bea
\frac{R^J_{\alpha'\alpha}(p)}
{4\mu i^{-l'_y+l_y+1} p^{l'_y+l_y+1}}
=\frac{1}{p^{l'_y+l_y}}
{}^{(-)}\la \Psi,(l'_y{\cal S}')J J_z|V^{PV}_n|\Psi,(l_y{\cal S})J J_z\ra^{(+)}
\eea
For the case of parity violation, we fix $l'_y=1$ and $l_y=0$.
To obtain the
observable parameters when neutron energies are larger than thermal ones
 (which correspond to zero energy limit for neutron spin rotation), one can use a simple extrapolation based on the above representation with a good accuracy up to  hundreds KeV.

The contributions to
parity violating matrix elements
$\frac{2}{\pi} \frac{1}{c_n}{\rm Im}\left[
\frac{R^J_{\alpha'\alpha}(p)}{4\mu p^{2}}\right]$
from different terms of parity violating potentials (see Table \ref{tbl:pvpotential}) are presented
in the Table \ref{tbl:matv:av18u9}. These matrix elements were calculated using strong AV18+UIX and weak DDH-II parity violating potentials for the case of
 low neutron energies (up to thermal ones). From this table, one can see that the main contribution to PV effects comes from $J=3/2$ channel for the ``best values'' of DDH coupling constants.

\begin{table}
\caption{\label{tbl:matv:av18u9}
Contributions to
$\frac{2}{\pi} {\rm Im}\left[
\frac{R^J_{1{\cal S}',0J}(p)}{4\mu p^{2}}\right]$
at very low energy in ${\rm fm}^2$ units.
We chose AV18+UIX as strong potential and
DDH-II as parity violating potential.
Matrix elements of $n=6,7$ are zero due to of isospin structure.
}
\begin{ruledtabular}
\begin{tabular}{rrrrr}
  n  & ${\cal S'}=\frac{1}{2},J=\frac{1}{2}$
     & ${\cal S'}=\frac{3}{2},J=\frac{1}{2}$
     & ${\cal S'}=\frac{1}{2},J=\frac{3}{2}$
     & ${\cal S'}=\frac{3}{2},J=\frac{3}{2}$ \\ \hline
  1&  $0.253\times 10^{+00} $ & $ 0.131\times 10^{+00} $ & $ -0.151\times 10^{-01} $ & $ -0.522\times 10^{+00} $ \\
  2 & $  -0.182\times 10^{-01} $ & $ -0.105\times 10^{-01} $ & $ 0.882\times 10^{-02} $ & $ 0.480\times 10^{-03} $ \\
  3&  $0.339\times 10^{-02} $ & $ 0.231\times 10^{-01} $ & $ -0.428\times 10^{-02} $ & $ -0.284\times 10^{-03} $ \\
  4&  $0.410\times 10^{-02} $ & $ -0.154\times 10^{-01} $ & $ 0.221\times 10^{-03} $ & $ 0.797\times 10^{-04} $ \\
  5&  $0.475\times 10^{-02} $ & $ -0.178\times 10^{-01} $ & $ 0.313\times 10^{-03} $ & $ 0.664\times 10^{-04} $ \\
  8&  $0.190\times 10^{-02} $ & $ 0.180\times 10^{-01} $ & $ -0.301\times 10^{-02} $ & $ -0.228\times 10^{-03} $ \\
  9&  $-0.562\times 10^{-02} $ & $ 0.960\times 10^{-02} $ & $ 0.107\times 10^{-02} $ & $ 0.278\times 10^{-04} $ \\
 10&  $0.388\times 10^{-02} $ & $ -0.146\times 10^{-01} $ & $ 0.209\times 10^{-03} $ & $ 0.755\times 10^{-04} $ \\
 11&  $0.453\times 10^{-02} $ & $ -0.170\times 10^{-01} $ & $ 0.298\times 10^{-03} $ & $ 0.631\times 10^{-04} $ \\
 12&  $0.452\times 10^{-02} $ & $ 0.165\times 10^{-03} $ & $ -0.223\times 10^{-03} $ & $ -0.105\times 10^{-01} $ \\
 13&  $0.725\times 10^{-02} $ & $ 0.113\times 10^{-02} $ & $ -0.377\times 10^{-03} $ & $ -0.175\times 10^{-01} $ \\
\end{tabular}
\end{ruledtabular}
\end{table}

Our results for the angle of neutron spin rotation for
 DDH, pionless EFT, and pionful EFT weak interaction potentials
with different sets of parameters
are summarized in Tables \ref{tbl:spinrot:DDH},\ref{tbl:spinrot:pionless}, and \ref{tbl:spinrot:pionful}. For these calculations, we used two types of strong interacting potentials:
Argonne two nucleon interaction AV18 and
inclusion of Urbana IX three nucleon interaction,
AV18+UIX.
One can see that these results practically
do not depend on a choice of the strong interaction potential.
Also, it is clear that the matrix element related to pion-exchange ($n=1$) is dominant for DDH potential,
slightly enhanced for pionfull potential,
and about equal to other ones for pionless potential.

\begin{table}
\caption{\label{tbl:spinrot:DDH}
Coefficients $I_n^{DDH}$ for
AV18 and AV18+UIX strong potentials,  and
DDH-I and DDH-II parameter
sets for parity violating potentials.
$I^{DDH}_{6,7}=0$.
}
\begin{ruledtabular}
\begin{tabular}{rrrrr}
$n$ & DDH-I/AV18 & DDH-I/AV18+UIX
    & DDH-II/AV18 & DDH-II/AV18+UIX \\
    \hline
  1& $ 0.612\times 10^{+02 } $& $  0.596\times 10^{+02 } $& $  0.616\times 10^{+02 } $& $ 0.600\times 10^{+02 }$\\
  2& $ 0.666\times 10^{+00 } $& $  0.726\times 10^{+00 } $& $  0.114\times 10^{+01 } $& $ 0.124\times 10^{+01}$\\
  3& $ -0.130\times 10^{+01 } $& $  -0.133\times 10^{+01 } $& $  -0.212\times 10^{+01 } $& $ -0.217\times 10^{+01}$\\
  4& $ 0.911\times 10^{+00 } $& $  0.934\times 10^{+00 } $& $  0.131\times 10^{+01 } $& $ 0.134\times 10^{+01}$\\
  5& $ 0.980\times 10^{+00 } $& $  0.992\times 10^{+00 } $& $  0.153\times 10^{+01 } $& $ 0.156\times 10^{+01}$\\
  8& $ -0.125\times 10^{+01 } $& $  -0.130\times 10^{+01 } $& $  -0.160\times 10^{+01 } $& $ -0.167\times 10^{+01}$\\
  9& $ -0.615\times 10^{+00 } $& $  -0.622\times 10^{+00 } $& $  -0.786\times 10^{+00 } $& $ -0.796\times 10^{+00}$\\
 10& $ 0.998\times 10^{+00 } $& $  0.102\times 10^{+01 } $& $  0.124\times 10^{+01 } $& $ 0.127\times 10^{+01}$\\
 11& $ 0.111\times 10^{+01 } $& $  0.113\times 10^{+01 } $& $  0.146\times 10^{+01 } $& $ 0.149\times 10^{+01}$\\
 12& $ 0.991\times 10^{+00 } $& $  0.983\times 10^{+00 } $& $  0.141\times 10^{+01 } $& $ 0.140\times 10^{+01}$\\
 13& $ 0.144\times 10^{+01 } $& $  0.144\times 10^{+01 } $& $  0.226\times 10^{+01 } $& $ 0.225\times 10^{+01}$\\
\end{tabular}
\end{ruledtabular}
\end{table}

\begin{table}
\caption{\label{tbl:spinrot:pionless}
Coefficients $I_n^{\not{\pi}}$ for
AV18 and AV18+UIX strong potentials, and
$\not{\pi}$EFT-I and $\not{\pi}$EFT-II
parameter sets for parity violating potentials.
$I^{\not{\pi}}_{2,3,5,6,7,10,11,12,13}=0$.
}
\begin{ruledtabular}
\begin{tabular}{crrrr}
$n$ & $\not{\pi}$EFT-I/AV18 & $\not{\pi}$EFT-I/AV18+UIX
    & $\not{\pi}$EFT-II/AV18 & $\not{\pi}$EFT-II/AV18+UIX \\
    \hline
1&   $0.616\times 10^{+02}$&   $0.600\times 10^{+02}$&   $0.969\times 10^{+00}$&  $0.969\times 10^{+00}$\\
4&   $0.606\times 10^{+02}$&   $0.588\times 10^{+02}$&   $0.499\times 10^{+00}$&  $0.515\times 10^{+00}$\\
8&   $-0.761\times 10^{+02}$&   $-0.757\times 10^{+02}$&   $-0.677\times 10^{+00}$&  $-0.708\times 10^{+00}$\\
9&   $-0.946\times 10^{+01}$&   $-0.662\times 10^{+01}$&   $-0.341\times 10^{+00}$&  $-0.348\times 10^{+00}$\\
\end{tabular}
\end{ruledtabular}
\end{table}

\begin{table}
\caption{\label{tbl:spinrot:pionful}
Coefficients $I_n^{\pi}$ for
AV18 and AV18+UIX strong potentials,  and
${\pi}$EFT-I and ${\pi}$EFT-II
parameter sets for parity violating potentials.
$I^{\pi}_{2,3,6,7,10,11,12}=0$.
}
\begin{ruledtabular}
\begin{tabular}{crrrr}
$n$ & ${\pi}$EFT-I/AV18 & ${\pi}$EFT-I/AV18+UIX
    & ${\pi}$EFT-II/AV18 & ${\pi}$EFT-II/AV18+UIX \\
    \hline
  1&$0.616\times 10^{+02}$&$0.600\times 10^{+02}$&$ 0.616\times 10^{+02}$&$ 0.600\times 10^{+02}$\\
  4&$0.152\times 10^{+01}$&$0.142\times 10^{+01}$&$ 0.549\times 10^{+00}$&$ 0.488\times 10^{+00}$\\
  5&$0.435\times 10^{+01}$&$0.185\times 10^{+01}$&$ 0.123\times 10^{+01}$&$ 0.664\times 10^{-01}$\\
  8&$-0.184\times 10^{+01}$&$-0.179\times 10^{+01}$&$ -0.782\times 10^{+00}$&$ -0.748\times 10^{+00}$\\
  9&$-0.820\times 10^{+00}$&$-0.730\times 10^{+00}$&$ -0.340\times 10^{+00}$&$ -0.288\times 10^{+00}$\\
 13&$0.226\times 10^{+02}$&$0.218\times 10^{+02}$&$ 0.970\times 10^{+01}$&$ 0.936\times 10^{+01}$\\
 14&$0.339\times 10^{+01}$&$0.333\times 10^{+01}$&$ 0.177\times 10^{+01}$&$ 0.174\times 10^{+01}$\\
 15&$0.654\times 10^{+02}$&$0.631\times 10^{+02}$&$ 0.273\times 10^{+02}$&$ 0.264\times 10^{+02}$\\
\end{tabular}
\end{ruledtabular}
\end{table}

The  neutron spin asymmetry $P$ was calculated for laboratory neutron energy  $E=15$ KeV. The results  are summarized in tables
\ref{tbl:asymm:DDH}, \ref{tbl:asymm:pionless}, and  \ref{tbl:asymm:pionful} for
 DDH, pionless EFT, and pionful EFT weak interaction potentials with different sets of parameters, correspondingly. These results provide a pattern similar to that of the results for the angle of neutron spin rotation.   The parameter
 $J_n$  in these tables is defined as
\bea
J_n\equiv \frac{1}{c_n}\frac{2}{\pi}
{\rm Re}\left[\frac{1}{4 \mu p^2}
\left(
R^{\frac{1}{2}}_{1\frac{1}{2},0\frac{1}{2}}
  -2\sqrt{2} R^{\frac{1}{2}}_{1\frac{3}{2},0\frac{1}{2}}
    +4R^{\frac{3}{2}}_{1\frac{1}{2},0\frac{3}{2}}
  -2\sqrt{5} R^{\frac{3}{2}}_{1\frac{3}{2},0\frac{3}{2}}
\right)
\right],
\eea
and  is related to the parameter $\tilde{I}_n$ in the expression  $P=\sum c_n \tilde{I}_n$ by
\bea
\tilde{I}_n
=\frac{\frac{2}{3}(2\pi \mu p^2) J_n}{
{\rm Re}\left[R^{\frac{1}{2}}_{0\frac{1}{2},0\frac{1}{2}}
    +2 R^{\frac{3}{2}}_{0\frac{3}{2},0\frac{3}{2}}\right]}
=\frac{8\pi^2\mu}{9}\frac{J_n}{\sigma_{tot}}
,
\eea
where
$\sigma_{tot}$ is the total $n-d$ cross section. The total cross section $\sigma_{tot}$ can be calculated, or one can use its known
 experimental value.

\begin{table}
\caption{\label{tbl:asymm:DDH}
Coefficients ${J}_n^{DDH}$ for
AV18 and AV18+UIX strong potentials, and
DDH-I and DDH-II parameter
sets for parity violating potentials
at $E=15$ KeV in the laboratory frame.
$J^{DDH}_{6,7}=0$.
}
\begin{ruledtabular}
\begin{tabular}{crrrr}
$n$ & DDH-I/AV18 & DDH-I/AV18+UIX
    & DDH-II/AV18 & DDH-II/AV18+UIX \\
    \hline
  1&$  0.253\times 10^{+00}$&$   0.253\times 10^{+00}$&$   0.254\times 10^{+00}$&$  0.254\times 10^{+00}$\\
  2&$  0.246\times 10^{-02}$&$   0.245\times 10^{-02}$&$   0.390\times 10^{-02}$&$  0.384\times 10^{-02}$\\
  3&$  -0.190\times 10^{-02}$&$   -0.147\times 10^{-02}$&$   -0.313\times 10^{-02}$&$  -0.243\times 10^{-02}$\\
  4&$  0.769\times 10^{-03}$&$   0.393\times 10^{-03}$&$   0.110\times 10^{-02}$&$  0.563\times 10^{-03}$\\
  5&$  0.846\times 10^{-03}$&$   0.442\times 10^{-03}$&$   0.132\times 10^{-02}$&$  0.689\times 10^{-03}$\\
  8&$  -0.176\times 10^{-02}$&$   -0.134\times 10^{-02}$&$   -0.228\times 10^{-02}$&$  -0.175\times 10^{-02}$\\
  9&$  -0.235\times 10^{-03}$&$   0.567\times 10^{-04}$&$   -0.259\times 10^{-03}$&$  0.118\times 10^{-03}$\\
 10&$  0.842\times 10^{-03}$&$   0.430\times 10^{-03}$&$   0.104\times 10^{-02}$&$  0.534\times 10^{-03}$\\
 11&$  0.957\times 10^{-03}$&$   0.500\times 10^{-03}$&$   0.126\times 10^{-02}$&$  0.657\times 10^{-03}$\\
 12&$  0.374\times 10^{-02}$&$   0.370\times 10^{-02}$&$   0.528\times 10^{-02}$&$  0.522\times 10^{-02}$\\
 13&$  0.563\times 10^{-02}$&$   0.559\times 10^{-02}$&$   0.874\times 10^{-02}$&$  0.868\times 10^{-02}$\\
\end{tabular}
\end{ruledtabular}
\end{table}

\begin{table}
\caption{\label{tbl:asymm:pionless}
Coefficients $J_n^{\not{\pi}}$ for
AV18 and AV18+UIX strong potentials, and
$\not{\pi}$EFT-I and $\not{\pi}$EFT-II
parameter sets for parity violating potentials.
$J^{\not{\pi}}_{2,3,5,6,7,10,11,12,13}=0$.
}
\begin{ruledtabular}
\begin{tabular}{crrrr}
$n$ & $\not{\pi}$EFT-I/AV18 & $\not{\pi}$EFT-I/AV18+UIX
    & $\not{\pi}$EFT-II/AV18 & $\not{\pi}$EFT-II/AV18+UIX \\
    \hline
1&$ 0.254\times 10^{+00}$&$ 0.254\times 10^{+00}$&$ 0.372\times 10^{-02}$&$ 0.369\times 10^{-02 }$\\
4&$ 0.503\times 10^{-01}$&$ 0.240\times 10^{-01}$&$ 0.421\times 10^{-03}$&$ 0.215\times 10^{-03 }$\\
8&$ -0.111\times 10^{+00}$&$ -0.854\times 10^{-01}$&$ -0.984\times 10^{-03}$&$ -0.763\times 10^{-03 }$\\
9&$ -0.241\times 10^{-02 }$&$ 0.338\times 10^{-02}$&$ -0.904\times 10^{-04}$&$ 0.750\times 10^{-04 }$\\
\end{tabular}
\end{ruledtabular}
\end{table}

\begin{table}
\caption{\label{tbl:asymm:pionful}
Coefficients $J_n^{\pi}$ for
AV18 and AV18+UIX strong potentials, and
${\pi}$EFT-I and ${\pi}$EFT-II
parameter sets for parity violating potentials.
$J^{\pi}_{2,3,6,7,10,11,12}=0$.
}
\begin{ruledtabular}
\begin{tabular}{crrrr}
$n$ & ${\pi}$EFT-I/AV18 & ${\pi}$EFT-I/AV18+UIX
    & ${\pi}$EFT-II/AV18 & ${\pi}$EFT-II/AV18+UIX \\
    \hline
  1&$ 0.254\times 10^{+00}$&$ 0.254\times 10^{+00}$&$ 0.254\times 10^{+00}$&$ 0.254\times 10^{+00}$\\
  4&$ 0.106\times 10^{-02}$&$ 0.352\times 10^{-03}$&$ 0.309\times 10^{-03}$&$ 0.333\times 10^{-04}$\\
  5&$ 0.741\times 10^{-02}$&$ 0.512\times 10^{-02}$&$ 0.292\times 10^{-02}$&$ 0.221\times 10^{-02}$\\
  8&$ -0.276\times 10^{-02}$&$ -0.212\times 10^{-02}$&$ -0.127\times 10^{-02}$&$ -0.100\times 10^{-02}$\\
  9&$ -0.148\times 10^{-03}$&$ 0.301\times 10^{-03}$&$ -0.278\times 10^{-04}$&$ 0.168\times 10^{-03}$\\
 13&$ 0.976\times 10^{-01}$&$ 0.981\times 10^{-01}$&$ 0.421\times 10^{-01}$&$ 0.423\times 10^{-01}$\\
 14&$ 0.137\times 10^{-01}$&$ 0.136\times 10^{-01}$&$ 0.714\times 10^{-02}$&$ 0.712\times 10^{-02}$\\
 15&$ 0.283\times 10^{+00}$&$ 0.284\times 10^{+00}$&$ 0.119\times 10^{+00}$&$ 0.120\times 10^{+00}$\\
\end{tabular}
\end{ruledtabular}
\end{table}


From the presented data, one can see that the results of our calculations are only slightly different  for the cases when we use  AV18 and AV18+UIX strong Hamiltonians.  This indicates  stability of the results with respect to the three nucleon
forces. Indeed,
by analyzing the DDH one-pion exchange matrix element (see Table \ref{tbl:matv:av18u9}), one can see that
for DDH-I with potentials AV18 and AV18+UIX, the contributions to the $I_{n=1}$ are $-0.180\times 10^{+01}$ and $-0.333\times 10^{+01}$  for doublet channel ($J=1/2$), and for the quartet channel ($J=3/2$) they are  $0.630\times 10^{+02}$ and $0.630\times 10^{+02}$, correspondingly.
The quartet channel is dominated by the repulsive and  long-range part of the strong interactions, but the doublet channel is defined by  attractive part. Therefore the quartet channel is less sensitive to the off-energy shell structure of the strong interactions compared to the doublet channel.
Then, due to the dominant contribution from the quartet channel, the net result
turns to be rather independent on the contribution from  three nucleon
forces. This fact demonstrates the independence of our results on models of strong interactions. However, further investigations with different strong interaction potentials are desirable.

It should be noted, that the dependence on cutoff parameters for the contributions from potentials with short  and middle range interactions, even though  it appears large, does not lead to cutoff dependence for the observable parameters. Indeed,  the renormalization of low energy constants would cancel
 those cutoff dependencies
by the cutoff dependencies of LECs.
Therefore, as a result, calculated PV
 observables are practically cutoff independent.

All these tables present information about contributions of different PV operators to PV effects, provided we know corresponding weak coupling constants. Then, to calculate parity violating effects, we can use either DDH potential or one of the considered EFT potentials. However, for the case of EFT potentials, we need to know a  set of
 LECs which cannot be calculated in the given theoretical framework but  must be obtained from a number of independent experiments. Unfortunately, currently available experimental data are not enough to define the LECs  with required  precision. Even for pionless EFT, the estimated LECs  \cite{Zhu:2004vw} have
 large uncertainties preventing us from predicting the values of PV effects.   For the
 pionful EFT, the situation with determination of LECs is even worse.
Therefore, it is impossible to make  reliable predictions for PV effects using EFT-type potentials at this time, and the only reasonable way to  estimate  magnitudes of PV effects is to use  the DDH potential. Taking into account  the difficulty of the systematic description of PV effects using ``standard'' DDH potentials (see discussions in the introduction), we  estimate PV effects
 using the DDH potential for different sets of weak coupling constants: both for the ``best value'' coupling constants and for two possible sets of the values of the coupling constants recently obtained by Bowman \cite{Bowman} from the fit of  reliable   existing experimental data
(see Table \ref{tbl:LEC:DDH}).
The results for these three sets of weak coupling constants are summarized in  Tables \ref{tbl:observables:DDH} and \ref{tbl:obs:asym:DDH} for the angle of spin rotation and for neutron spin asymmetry, correspondingly.
One can see that in contrast to the fact that the one-pion exchange
dominates in the DDH-``best" coupling parameter set, the
rho meson exchange dominates in the case of
Bowman's coupling parameter set.  One can see that the angle of neutron spin rotation has almost the same magnitude for all three sets of parameters, but it has  opposite signs for the   ``best value'' set and for the Bowman's fits. The neutron spin asymmetry does not only have opposite signs but also essentially different values for these two choices of parameters. This allows one to choose between two possible sets of DDH parameters and, as a consequence, to test the dominance of pion-meson contribution in PV effects in $n-d$ scattering.

\begin{table}
\caption{\label{tbl:LEC:DDH}
DDH PV coupling constants in units of $10^{-7}$.
Strong couplings are
$\frac{g^2_\pi}{4\pi}=13.9$,
$\frac{g^2_\rho}{4\pi}=0.84$,
$\frac{g^2_\omega}{4\pi}=20$,
$\kappa_\rho=3.7$, and
$\kappa_\omega=0$, $h'_\rho$ contribution is neglected.
4-paramter fir and 3-parameter fit uses the same $h_\rho^1$
and $h_\omega^1$ with DDH `best'.
}
\begin{ruledtabular}
\begin{tabular}{cccc}
DDH Coupling& DDH `best' & 4-parameter fit\cite{Bowman}
                         & 3-parameter fit\cite{Bowman} \\
\hline
$h^1_\pi$ &  $+4.56$        &  $-0.456$&  $-0.5$ \\
$h_\rho^0$ & $-11.4$        &  $-43.3$  & $-33$ \\
$h_\rho^2$ & $-9.5$         & $37.1$   & $41$ \\
$h_\omega^0$&  $-1.9$       & $13.7$  & $0$ \\
$h_\rho^1$ &   $-0.19$    &  $-0.19$ & $-0.19$ \\
$h_\omega^1$ & $-1.14$    &  $-1.14$ & $-1.14$ \\
\end{tabular}
\end{ruledtabular}
\end{table}

\begin{table}
\caption{\label{tbl:observables:DDH}
Neutron spin rotation in
$10^{-7}\mbox{ rad-cm}^{-1}$ for the case of DDH-II potential
with AV18+UIX strong potential
  for
 a liquid deuteron density
$N=0.4\times 10^{23}$ atoms per $cm^{3}$.
}
\begin{ruledtabular}
\begin{tabular}{crrr}
  & DDH 'best' &4-parameter fit\cite{Bowman} & 3-parameter fit\cite{Bowman} \\
\hline
  1&$  0.108\times 10^{+00}$&$   -0.108\times 10^{-01}$&$    -0.118\times 10^{-01}$\\
  2&$  0.386\times 10^{-02}$&$    0.147\times 10^{-01}$&$     0.112\times 10^{-01}$\\
  3&$ -0.317\times 10^{-01}$&$   -0.120\times 10^{+00}$&$    -0.918\times 10^{-01}$\\
  4&$  0.349\times 10^{-04}$&$    0.349\times 10^{-04}$&$     0.349\times 10^{-04}$\\
  5&$  0.150\times 10^{-03}$&$    0.150\times 10^{-03}$&$     0.150\times 10^{-03}$\\
  8&$ -0.423\times 10^{-02}$&$    0.305\times 10^{-01}$&$     0.000\times 10^{+00}$\\
  9&$ -0.202\times 10^{-02}$&$    0.146\times 10^{-01}$&$     0.000\times 10^{+00}$\\
 10&$  0.967\times 10^{-03}$&$    0.967\times 10^{-03}$&$     0.967\times 10^{-03}$\\
 11&$  0.113\times 10^{-02}$&$    0.113\times 10^{-02}$&$     0.113\times 10^{-02}$\\
 12&$  0.102\times 10^{-02}$&$    0.102\times 10^{-02}$&$     0.102\times 10^{-02}$\\
total&$  0.768\times 10^{-01}$&$   -0.682\times 10^{-01}$&$    -0.891\times 10^{-01}$\\
\end{tabular}
\end{ruledtabular}
\end{table}

\begin{table}
\caption{\label{tbl:obs:asym:DDH}
Neutron spin asymmetry for the case of DDH-II potential
with AV18+UIX strong potential (the total cross section
$\sigma_{tot}=3.35\mbox{ b}$ at $E=15$ KeV ).
}
\begin{ruledtabular}
\begin{tabular}{crrr}
  & DDH 'best' &4-parameter fit\cite{Bowman} & 3-parameter fit\cite{Bowman} \\
\hline
  1&$  0.947\times 10^{-08}$&$    -0.947\times 10^{-09}$&$    -0.104\times 10^{-08}$\\
  2&$  0.248\times 10^{-09}$&$    0.943\times 10^{-09}$&$    0.719\times 10^{-09}$\\
  3&$  -0.740\times 10^{-09}$&$    -0.281\times 10^{-08}$&$    -0.214\times 10^{-08}$\\
  4&$  0.304\times 10^{-12}$&$    0.304\times 10^{-12}$&$    0.304\times 10^{-12}$\\
  5&$  0.138\times 10^{-11}$&$    0.138\times 10^{-11}$&$    0.138\times 10^{-11}$\\
  8&$  -0.922\times 10^{-10}$&$    0.665\times 10^{-09}$&$    0.000\times 10^{+00}$\\
  9&$  0.620\times 10^{-11}$&$    -0.447\times 10^{-10}$&$    -0.000\times 10^{+00}$\\
 10&$  0.843\times 10^{-11}$&$    0.843\times 10^{-11}$&$    0.843\times 10^{-11}$\\
 11&$  0.104\times 10^{-10}$&$    0.104\times 10^{-10}$&$    0.104\times 10^{-10}$\\
 12&$  0.797\times 10^{-10}$&$    0.797\times 10^{-10}$&$    0.797\times 10^{-10}$\\
 \hline
total&$  0.899\times 10^{-08}$&$   -0.209\times 10^{-08}$&$   -0.236\times 10^{-08}$\\
\end{tabular}
\end{ruledtabular}
\end{table}

Finally, we would like to mention that
 our results are quite different from  the results obtained in paper
 \cite{Schiavilla:2008ic}. For example, in paper \cite{Schiavilla:2008ic}, the values of
$I_n$ for $J=\frac{1}{2}$ and $J=\frac{3}{2}$
have the same signs for operator  with $n=1$, but our results show opposite signs for these matrix elements.
Another discrepancy  is related to the systematic difference between the values of matrix elements calculated \cite{Schiavilla:2008ic} for AV18 and AV18+UIX potentials, which indicates a large wave function difference for
 AV18 and AV18+UIX potentials. Contrary to those, our results show that these matrix elements are
insensitive to the presence of the three nucleon force.
\footnote{ We  thank  R. Schiavilla and M. Viviani
for discussions which clarified that the reason for theses discrepancies is related to numerical errors in paper \cite{Schiavilla:2008ic}. }

\section{Conclusion}

We have calculated parity violating  angle of neutron spin rotation and  asymmetry in transmission of neutrons with opposite helicities  for low energy  neutron deuteron scattering. Using Distorted Wave Born Approximation for weak interactions with realistic three-nucleon wave functions from Faddeev equations in configuration space, we have parameterized PV observables in terms of matrix elements presented in the DDH weak potential and in weak potentials derived from pionless and pionful EFTs. It is shown that our results practically do not depend on the choice for strong interaction potentials and on cutoff parameters.

Based on the given analysis, one can see that for  DDH potential, the dominant contribution to observable PV effects comes from the pion-exchange matrix element with $n=1$.  However, for pionless EFT potential, all types of matrix elements contribute almost equally, and for pionful EFT potential the pion-exchange matrix element is sightly enhanced as compared to the other ones.
Therefore, it would be interesting to compare the estimation of observable PV effects using appropriate LECs and coupling constants for DDH. Unfortunately, due to insufficient data for LECs this is impossible at this time.
However, a comparison  of PV effects for two different sets of coupling constants  shows that  $n-d$ scattering experimental results can be used to distinguish between different sets of DDH coupling constants and to help in clarification of the issue about the importance of the contribution of pion-exchange weak potential.

\appendix*
\section{Explicit results of angular part of Matrix elements}
Explicit values of  matrix elements of iso-spin operators for two-body states are
\bea
\la T' T'_z|\tau_1\cdot\tau_2| T T_z\ra
&=&\delta_{T'_z,T_z}\delta_{T',T}[1\delta_{T,1}-3\delta_{T,0}],\no
\la T' T'_z|(\tau_1+\tau_2)^z|T T_z\ra
&=&\delta_{T' 1}\delta_{T 1}\delta_{T'_z T_z}[2 T_z],\no
\la T' T'_z|(\tau_1-\tau_2)^z|T T_z\ra
&=&\delta_{T',T\pm 1}\delta_{T_z,T'_z}\delta_{T_z,0}[2],\no
\la T' T'_z|i(\tau_1\times\tau_2)^z|T T_z\ra
&=&\delta_{T'_z,T_z}\delta_{T_z,0}\delta_{T',T\pm 1}[\pm 2],\no
\la T' T'_z|{\cal T}^z_{12}|T T_z\ra
&=&\delta_{T',1}\delta_{T,1}\delta_{T'_z, T_z}
[2\delta_{T_z,1}-4\delta_{T_z,0}+2\delta_{T_z,-1}],
\eea
and matrix elements of orbital and spin operators
for two-body states $|(l_x s_x)j_x j_x^z\rangle$ are
\bea
& &\la (j_x\pm 1, 1) j_x j^z_x| (\vs_1+ \vs_2)\cdot{\hat x}| (j_x,1) j_x j^z_x\ra\no
&=&\la (j_x, 1) j_x j^z_x| (\vs_1+ \vs_2)\cdot{\hat x}
 |(j_x\pm 1,1) j_x j^z_x\ra \no
&=&-2 \sqrt{\frac{j_x+1/2\mp 1/2}{2j_x+1}}
\eea
\bea
& &\la (j_x\pm 1, 1)j_x j_x^z|(\vs_1-\vs_2)\cdot\hat{x}|(j_x,0)j_x j_x^z\ra\no
&=&\la (j_x, 0)j_x j_x^z|(\vs_1-\vs_2)\cdot\hat{x}|(j_x\pm 1,1)j_x j_x^z\ra\no
&=&\mp 2\sqrt{\frac{j_x+1/2\pm 1/2}{2j_x+1}}
\eea
\bea
& &\la (j_x, 0)j_x j_x^z|i(\vs_1\times\vs_2)\cdot\hat{x}|(j_x\pm 1,1)j_x j_x^z\ra\no
&=&(-)\la (j_x\pm 1, 1)j_x j_x^z|i(\vs_1\times\vs_2)\cdot\hat{x}|(j_x,0)j_x j_x^z\ra \no
&=& \pm 2  \sqrt{\frac{j_x+1/2\pm 1/2}{2j_x+1}}
\eea
\bea
\la (j_x\pm1 1) j_x j^z_x|(\vs_1+\vs_2)\cdot \hat{\nabla}_{\Omega_x}
    |(j_x 1) j_x j^z_x\ra
&=& \pm 2  \frac{(j_x+1/2\mp 1/2) \sqrt{j_x+1/2\mp 1/2}}{\sqrt{2j_x+1}}
\no
\la (j_x 1) j_x j^z_x|(\vs_1+\vs_2)\cdot \hat{\nabla}_{\Omega_x}
    |(j_x\pm 1 1) j_x j^z_x\ra
&=& \mp 2  \frac{(j_x+1/2\pm 3/2) \sqrt{j_x+1/2\mp 1/2}}{\sqrt{2j_x+1}}
\eea
\bea
\la (j_x\pm1 1) j_x j^z_x|(\vs_1-\vs_2)\cdot \hat{\nabla}_{\Omega_x}
    |(j_x 0) j_x j^z_x\ra
&=& 2  \frac{(j_x+1/2\mp 1/2) \sqrt{j_x+1/2\pm 1/2}}{\sqrt{2j_x+1}}
\no
\la (j_x 0) j_x j^z_x|(\vs_1-\vs_2)\cdot \hat{\nabla}_{\Omega_x}
    |(j_x\pm1 1) j_x j^z_x\ra
&=& -2  \frac{(j_x+1/2\pm 3/2) \sqrt{j_x+1/2\pm 1/2}}{\sqrt{2j_x+1}}
\eea


\begin{acknowledgments}
This work was supported by the DOE grants no. DE-FG02-09ER41621.
\end{acknowledgments}

\bibliography{ParityViolation}

\end{document}